\documentclass[trackchanges]{aastex7}
\usepackage{subfig}   
\usepackage{siunitx}      
\usepackage{amsmath}

\usepackage{booktabs}

\begin{document}

\title{Micrometeoroid Impacts: Dual Pathways for Iron Reduction and Oxidation on Lunar and Asteroidal Surfaces}
 
\author[orcid=0000-0002-8624-1264]{Ziyu Huang}
\affiliation{Daniel Guggenheim School of Aerospace Engineering, Georgia Institute of Technology, 620 Cherry
Street, Atlanta, GA 30332, USA.}
\email[show]{zyuhuang@gatech.edu}

\author[orcid=0000-0002-1821-5689	]{Masatoshi Hirabayashi}
\affiliation{Daniel Guggenheim School of Aerospace Engineering, Georgia Institute of Technology, 620 Cherry
Street, Atlanta, GA 30332, USA.}
\email[]{}

\author[orcid=0000-0002-2422-4506]{Thomas M. Orlando}
\affiliation{School of Chemistry and Biochemistry, Georgia Institute of Technology, 620 Cherry
Street, Atlanta, GA 30332, USA.}
\affiliation{School of Physics, Georgia Institute of Technology, 620 Cherry
Street, Atlanta, GA 30332, USA.}

\email[]{}

\begin{abstract}

Nanophase metallic iron (npFe$^0$) is a key indicator of space weathering on the lunar surface, primarily attributed to solar wind irradiation and micrometeoroid impacts. Recent discoveries of hematite (Fe$_2$O$_3$), a highly oxidized form of iron, in the lunar polar regions challenge the prevailing understanding of the Moon's reducing environment. This study, using ReaxFF molecular dynamics simulations of micrometeoroid impacts on fayalite (Fe$_2$SiO$_4$), investigates the atomistic mechanisms leading to both reduced and oxidized iron species. 
Our simulations reveals that the high-temperature and pressure conditions at the impact crater surface produces a reduced iron environment while providing a transient oxygen-rich environment in the expanding plume.
Our findings bridge previously disparate observations—linking impact-driven npFe$^0$ formation to the puzzling presence of oxidized iron phases on the Moon, completing the observed strong dichotomous distribution of hematite between the nearside and farside of the Moon. 
These findings highlight that micrometeoroid impacts, by simultaneously generating spatially distinct redox environments, provide a formation mechanism that reconciles the ubiquitous identification of nanophase metallic iron (npFe$^0$) in returned lunar samples with Fe$^{3+}$ signatures detected by remote sensing. This underscores the dynamic nature of space weathering processes. For a more nuanced understanding of regolith evolution, we should also consider the presence of different generations or types of npFe$^0$, such as those formed from solar wind reduction versus impact disproportionation.

\end{abstract}
 


\section{Introduction} 
   
Nanophase metallic iron (npFe$^0$), typically ranging from a few to tens of nanometers, is a well-established and critical component of mature lunar regolith~\citep{Hapke2001,1993Sci...261.1305K,2000M&PS...35.1101P}. Its accumulation profoundly influences the optical properties of the Moon, leading to characteristic spectral reddening and darkening, and it serves as a primary indicator of the extent of space weathering and soil maturity~\citep{Pieters2016}. Consequently, a thorough understanding of the formation mechanisms of npFe$^0$ is essential for the accurate interpretation of remote sensing data from the Moon and other airless bodies, as well as for deciphering the complex alteration histories recorded in returned lunar samples and meteorites.

For decades, investigations into the origins of npFe$^0$ have predominantly centered on two main processes: solar wind irradiation and micrometeoroid impact-induced vaporization with subsequent condensation~\citep{Pieters2016}. Solar wind mechanisms involve the implantation of hydrogen ions, which can reduce Fe-bearing minerals like olivine, pyroxene, and ilmenite to Fe$^0$, or sputtering, which may preferentially remove oxygen, thereby enriching iron on grain surfaces~\citep{Hapke2001}. Micrometeoroid impacts, conversely, deliver intense, localized energy upon collision, causing melting and vaporization of regolith. The iron from these vaporized silicates is thought to recondense on grain surfaces as npFe$^0$-rich rims or become incorporated into agglutinitic glasses~\citep{Noble2007}.

Recent analyses of Chang'e-5 lunar samples have further illuminated impact-driven disproportionation ($3\text{Fe}^{2+} \rightarrow \text{Fe}^0 + 2\text{Fe}^{3+}$) as another significant pathway for npFe$^0$ formation~\citep{2022NatAs...6.1156L}. Simultaneously, the surprising detection of hematite (Fe$_2$O$_3$), a highly oxidized form of iron, in lunar polar regions by the Moon Mineralogy Mapper (M3) has challenged our understanding of lunar surface chemistry~\citep{Li2020}. Hematite's presence is perplexing as the Moon's surface is generally considered a reducing environment, lacking the significant free oxygen or liquid water typically associated with its formation on Earth, prompting new questions about localized oxidation processes~\citep{Li2020}.

The formation of lunar hematite has been hypothesized to involve several factors, including the potential role of oxygen ions from Earth's upper atmosphere delivered via the magnetotail, or interaction with water ice in permanently shadowed regions, particularly when exposed to solar wind or micrometeoroid bombardment~\citep{Li2020,Head2020}. While the Earth's wind can deliver oxygen, micrometeoroid impacts themselves provide the energy and transient conditions that could facilitate oxidative reactions. This presents an intriguing dichotomy: impacts are known to produce reduced iron, yet they might also contribute, under specific circumstances, to the formation of oxidized phases like hematite, possibly through complex interactions within the impact plume~\citep{Li2020,Stopar2018}.

 While existing models explain many space weathering observations, the precise atomistic pathways leading to the reduced npFe$^0$ and oxidized iron phases like hematite from Fe$^{2+}$, particularly those directly driven by the dynamic conditions within an impact event, remain less understood. This manuscript employs a novel approach to examine these divergent, impact-driven redox pathways in detail. We investigate how the extreme conditions within the impact crater itself---specifically, high temperatures and pressures on Fe-bearing minerals---predominantly foster reducing conditions favorable to the npFe$^0$ formation. Concurrently, we explore how the rapidly expanding and cooling impact plume, through potential interactions with tenuous ambient oxygen sources or via differential volatilization and recondensation, can create localized, transient oxygen-rich zones conducive to forming higher oxidation state iron species. By detailing how intense energy deposition and subsequent plume dynamics can simultaneously generate these spatially distinct redox environments, this work aims to reconcile seemingly contradictory observations of lunar surface alteration and advance understanding of planetary surface evolution.

\section{Methods}

\subsection{ReaxFF Molecular Dynamics}

To investigate the atomistic mechanisms of nanophase iron formation under hypervelocity impact conditions, we employed molecular dynamics (MD) simulations utilizing the ReaxFF reactive force field~\citep{vanDuin2001,Chenoweth2008}. ReaxFF is particularly well-suited for this study as it dynamically models chemical bond formation and breakage by calculating bond orders between atoms at each time step. Previous studies have used ReaxFF to understand lunar volatile cycle \cite{2021GeoRL..4893509H,2022GeoRL..4999333H,huang2021molecular,georgiou2025effect} This capability enables the explicit simulation of complex chemical reactions, phase transitions, and material decomposition that occur in the extreme temperature and pressure environments generated by micrometeoroid impacts, which are inaccessible to traditional non-reactive force fields. The ReaxFF formalism includes terms for covalent, Coulombic, and van der Waals interactions, with parameters derived from quantum mechanical calculations, ensuring a reasonable approximation of interatomic forces across a wide range of coordination environments and chemical states. The use of a charge equilibration scheme (e.g., QEq) within ReaxFF allows for dynamic charge distribution, which is critical for modeling redox reactions and the formation of both metallic and oxidized iron species~\citep{Nistor2006}. The interactions between {\color{black}atoms} are described using a reactive force field (ReaxFF) potential \citep{vanDuin2001}, which computes the total energy as a sum of multiple bonded and non-bonded contributions:

\begin{equation}
\begin{aligned}
E_{\text{ReaxFF}}(\{\vec{r}_{ij}\}, & \{\vec{r}_{ijk}\}, \{\vec{r}_{ijkl}\}, \{q_{i}\}, \{BO_{ij}\}) = E_{\text{bond}} + E_{\text{lp}} + E_{\text{over}} + E_{\text{under}} + \\
& E_{\text{val}} + E_{\text{pen}} + E_{\text{coa}} + E_{\text{tors}} + E_{\text{conj}} + E_{\text{hbond}} + E_{\text{vdWaals}} + E_{\text{Coulomb}}
\end{aligned}
\end{equation}

\noindent 
where the total energy $E_{ReaxFF}$ is a function of the inter-atomic distance between an atomic pair, $\vec{r}_{ij}$, triplets, $\vec{r}_{ijk}$, and quadruplets, $\vec{r}_{ijkl}$, as well as atomic charges $q_i$ and bond orders $BO_{ij}$ between an atomic pair. The valence interactions include the bonding energy $E_{bond}$, lone-pair energy $E_{lp}$, overcoordination energy $E_{over}$, undercoordination energy $E_{under}$, valence-angle energy $E_{val}$, penalty energy $E_{pen}$, three-body conjugation energy $E_{coa}$, torsion-angle energy $E_{tors}$, 4-body conjugation energy $E_{conj}$, and hydrogen bonding energy $E_{hbond}$. The noncovalent interactions comprise van der Waals energy $E_{vdWaals}$ and Coulomb energy $E_{Coulomb}$, which are screened by a taper function. 

The simulations were performed using the Fe/Si/O ReaxFF force field parameters developed and validated in previous studies for modeling silicate systems and iron oxides~\citep{Aryanpour2010,Buehler2007}. These parameters have been demonstrated to accurately reproduce material properties such as bond energies, reaction enthalpies, equation of state for relevant phases (e.g., fayalite, metallic iron, iron oxides like FeO, Fe$_3$O$_4$, and Fe$_2$O$_3$), and surface reaction dynamics, making them suitable for capturing the complex chemistry anticipated in our impact scenarios. The fidelity of the force field in describing iron oxidation states and the interaction between iron, silicon, and oxygen is paramount for addressing our central hypothesis concerning spatially distinct redox environments.

\subsection{Simulation Setup and Parameters}

\begin{figure}[!h]
    \centering
    \includegraphics[width=0.8\textwidth]{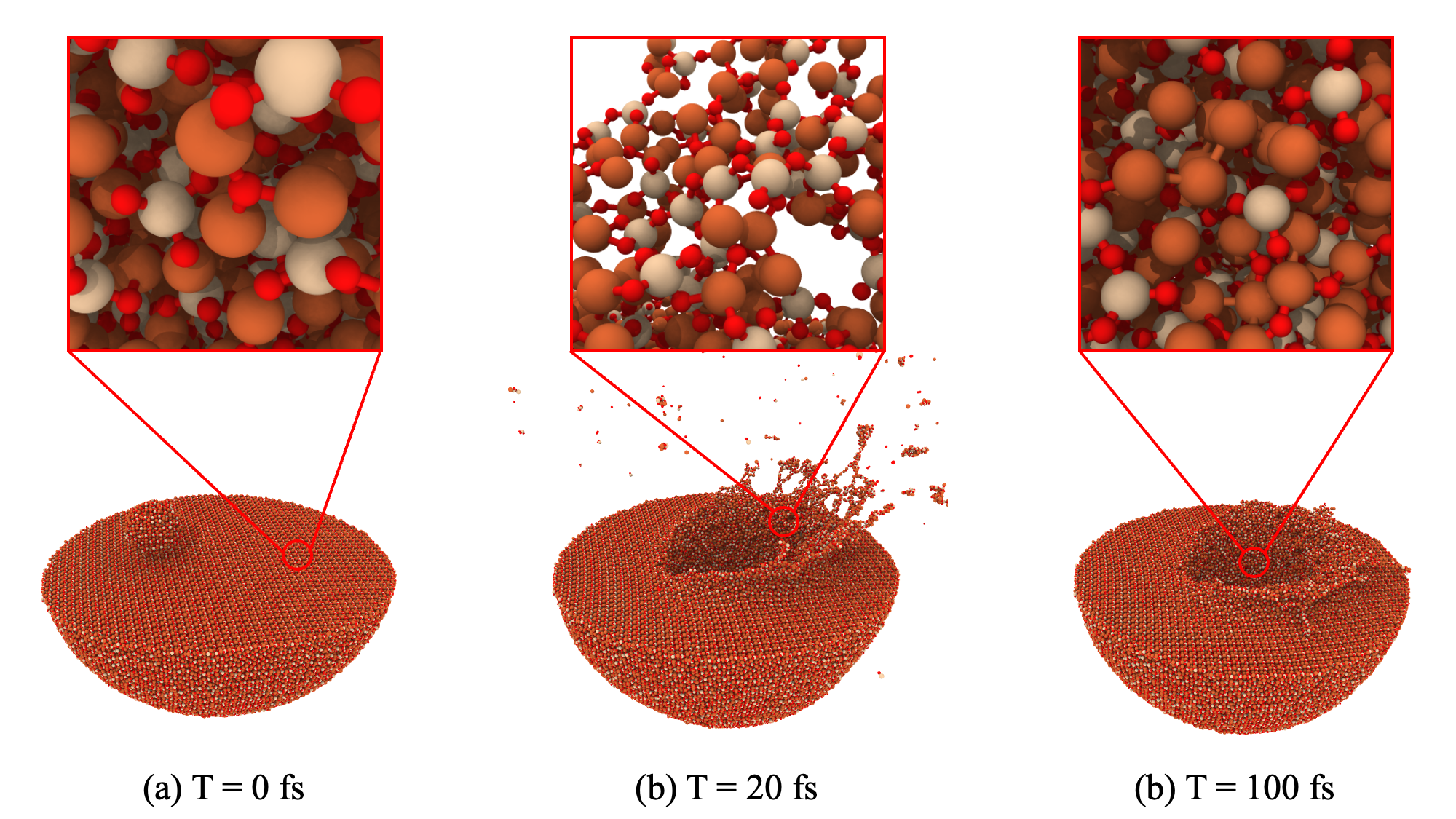}%
    \caption{Snapshots of the micrometeoroid impact simulation : (a) T = 0 fs – initial configuration, we use Fe$_2$SiO$_4$ unit cell to make the impactor and the substrate. (b) T = 20 fs – plume expansion, the developing amorphous structure is formed due to the impact. (c) T = 100 fs – post‐impact steady state, showing the local impact site and emergence of Fe–Fe bonds.}
    \label{fig:Setup}
\end{figure}

The target material was modeled as a crystalline fayalite (Fe$_2$SiO$_4$) half hemisphere with dimensions of approximately 24 nm, containing roughly 316,183 atoms, shown in Figure~\ref{fig:Setup}. Fayalite was chosen as a representative Fe-rich silicate mineral commonly found in lunar regolith~\citep{Papike1998}. The target slab was periodic in the x and y lateral directions, while the top surface (positive $z$-direction) was free to allow for ejecta plume development. Prior to impact, the fayalite substrate was equilibrated at 300 K using an NVT (canonical) ensemble to maintain number of atoms (N), volume of the system (V) and temperature (T)  for 20 ps to achieve thermal stability.

A spherical impactor, composed of crystalline fayalite (Fe$_2$SiO$_4$) with a diameter of \SI{5}{\nano\meter} (containing \num{2933} atoms), was utilized to simulate a micrometeoroid. 
{\color{black}
Although the impactor used in this study is only 5 nm—smaller than typical micrometeoroids—similar sub-micrometer craters have been observed in Chang’e-5 samples (Zeng et al., 2023; Gu et al., 2023). Redox differentiation arises from preferential oxygen ejection—a mechanism governed by atomistic-scale kinetics and expected to persist at larger scales despite quantitative differences in crater size or plume duration.
}
The impactor was initialized \SI{10}{\angstrom} (\SI{1}{\nano\meter}) above the fayalite target surface. Hypervelocity impact simulations were conducted with an impact velocity of \SI{12}{\kilo\meter\per\second}, directed at an incidence angle of \SI{45}{\degree} with respect to the surface normal. The defined impact speed comes from an average meteoroid impact speed on the lunar surface~\citep{Grun2011}, while the 45$^\circ$ incidence angle represents an average sense of meteoroid impact directions over a given hemispheric space \citep{melosh1989impact}. The impact simulations themselves were performed under the NVE (microcanonical) ensemble which maintains the number of atoms (N), volume of the system (V) and total energy (E) to conserve total system energy during the highly dynamic collision and subsequent plume expansion phase. Each impact and plume expansion simulation was run for a total duration of \SI{100}{\pico\second}, allowing for the observation of initial crater formation, ejecta evolution, and rapid chemical transformations.
{\color{black} The simulation duration was sufficient to capture the oxidation states of clusters sampled from the fully expanded plume (i.e., no collisions) and after surface temperature and pressure had stabilized and chemical reactions were no longer occurring. }
While longer simulations may reveal further evolution, the reported values capture the immediate redox state following the impact.
All MD simulations were performed using the LAMMPS (Large-scale Atomic/Molecular Massively Parallel Simulator) software package~\citep{Plimpton1995}.

\subsection{Analysis of Impact Ejecta}

Figure~\ref{fig:Setup}(b) shows the ejecta formed after the impact. Following the impact simulations, detailed post-processing analyses were conducted to characterize the resulting chemical and physical alterations. Atomic trajectories, temperature, pressure, and local atomic structures were recorded throughout the simulations. The primary analytical focus was on identifying and characterizing clusters of atoms, both within the modified crater region and in the ejecta plume. The ejecta plume was defined as all atoms located above the original target surface ($z>$  \SI{210}{\angstrom}) at a given time post-impact.

To characterize the formation and evolution of atomic aggregation, cluster analysis was performed utilizing a Density-Based Spatial Clustering of Applications with Noise (DBSCAN)-like algorithm \citep{schubert2017dbscan}. This approach was adapted to incorporate bond-based criteria combined with spatial proximity. Specifically, atoms were considered part of the same cluster if they were within a specified cutoff distance (e.g., $2.400$ Å for Fe-Fe, $2.112$ Å for Fe-O, $2.172$ Å for Si-O), consistent with typical bond lengths in the expected phases. A minimum cluster size of 2 atoms was typically imposed, analogous to the \texttt{minPts} parameter in DBSCAN, to filter out transient atomic associations and focus on more stable agglomerates. This analysis allowed for the identification of nascent nanophase metallic iron (npFe$^0$) clusters, iron oxide clusters (e.g., Fe$_x$O$_y$), and various silicate fragments.

To investigate our hypothesis of spatially distinct redox environments, we analyzed the elemental composition (relative atomic percentages of Fe, Si, and O) exclusively from atomic clusters sampled within the ejecta plume. For each identified cluster, particularly those rich in iron, the Fe:Si:O ratio was calculated to infer its oxidation state. In these calculations, we assumed fixed oxidation states for silicon as Si(+4) and oxygen as O(-2). This approach is justified because iron, as a transition metal, exhibits a more variable oxidation chemistry compared to silicon and oxygen, which tend to maintain their characteristic valences in common geological contexts. Therefore, charge balance within the cluster are closely related to the change of iron oxidation state.

{\color{black}
Clusters were classified as reduced if the average oxidation state of Fe was below +2, and as oxidized if it exceeded +2.
}
This comparative analysis of cluster compositions from the plume versus the crater region is designed to elucidate the differing chemical pathways active during and immediately after the impact event.

\section{Results and Discussions}

\begin{figure}[!h]
    \centering
    \includegraphics[width=1.0\textwidth]{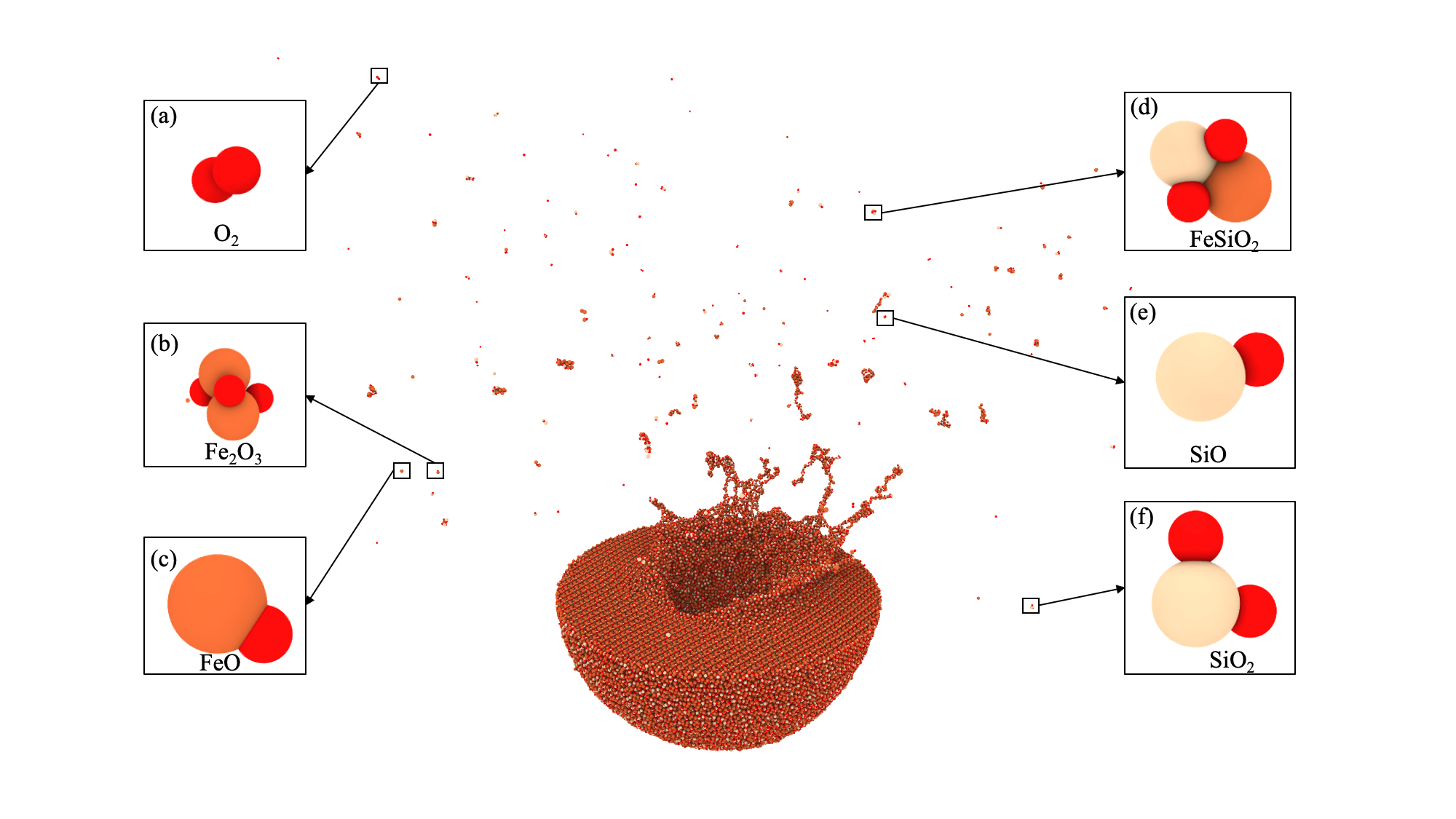}%
    \caption{A comprehensive view of the system during the ejecta expansion phase following micrometeoroid impact. Panels (a)–(f) show the most frequent chemical species identified by cluster analysis: (a) O$_2$, (b) Fe$_2$O$_3$, (c) FeO, (d) FeSiO$_2$, (e) SiO, and (f) SiO$_2$. The corresponding cluster counts for each species are plotted in Figure~\ref{fig:VDF_ejecta}.}
    \label{fig:Ejecta_Comp}
\end{figure}

The conditions at the impact site promote significant chemical alteration, with our findings on iron reduction aligning closely with recent work by \cite{shoji2025reactive}, who also documented impact-induced reduction of iron. Beyond iron, the chemical dynamics are notably complex, involving both reduction and oxidation reactions that occur simultaneously in different regions of the system. These competing processes reflect the interplay between localized reduction near the impact interface---where metallic Fe species can form---and regions in the surrounding matrix or ejecta plume, where oxidation may be favored.

{\color{black}
A notable outcome of our analysis is the elemental differentiation between the impact site and the ejected material, which we quantified by defining a hemispherical region with a radius of 80~\AA{} centered at the simulation box to encompass both the impact crater and the ejecta. By examining the elemental composition of the ejected material (Fe: 307 atoms, Si: 153 atoms, O: 649 atoms) and comparing it to the initial stoichiometry of the Fe$_2$SiO$_4$ target, we find that the ejecta is markedly enriched in oxygen relative to iron and silicon. As shown in Table~\ref{tab:elemental_composition}, the ejecta is characterized by a lower Fe/(O+Si) ratio compared to both the original target and the impact site, we find values of 0.4103 for the impact site and 0.3828 for the ejecta.  
This preferential ejection of oxygen supports our main conclusion that micrometeoroid impacts favor the release of lighter elements, likely due to oxygen's lower atomic mass and greater mobility, making it more susceptible to sputtering and ejection. Overall, our results demonstrate that disproportionation reactions at the impact site lead to both localized reduction and an oxygen-rich ejecta, resulting in an oxygen-poor surface after the impact.

}

\begin{table}[!h]
\centering
\begin{tabular}{lcccr}
\hline
\textbf{Region} & \textbf{Fe} & \textbf{Si} & \textbf{O} & \multicolumn{1}{l}{\textbf{Fe/(Si+O)}} \\ \hline
Setup           & 91176       & 45588       & 182352     & 0.4000                                      \\
Impact Site     & 18313       & 8947        & 35687      & 0.4103                             \\
Ejecta          & 307         & 153         & 649        & 0.3828                             \\ \hline
\end{tabular}
\caption{Comparison of elemental composition at the impact site and in the ejecta.}
\label{tab:elemental_composition}
\end{table}

\subsection{Composition of Ejecta Clusters}

\begin{figure}[!h]
    \centering

    \includegraphics[width=1.0\textwidth]{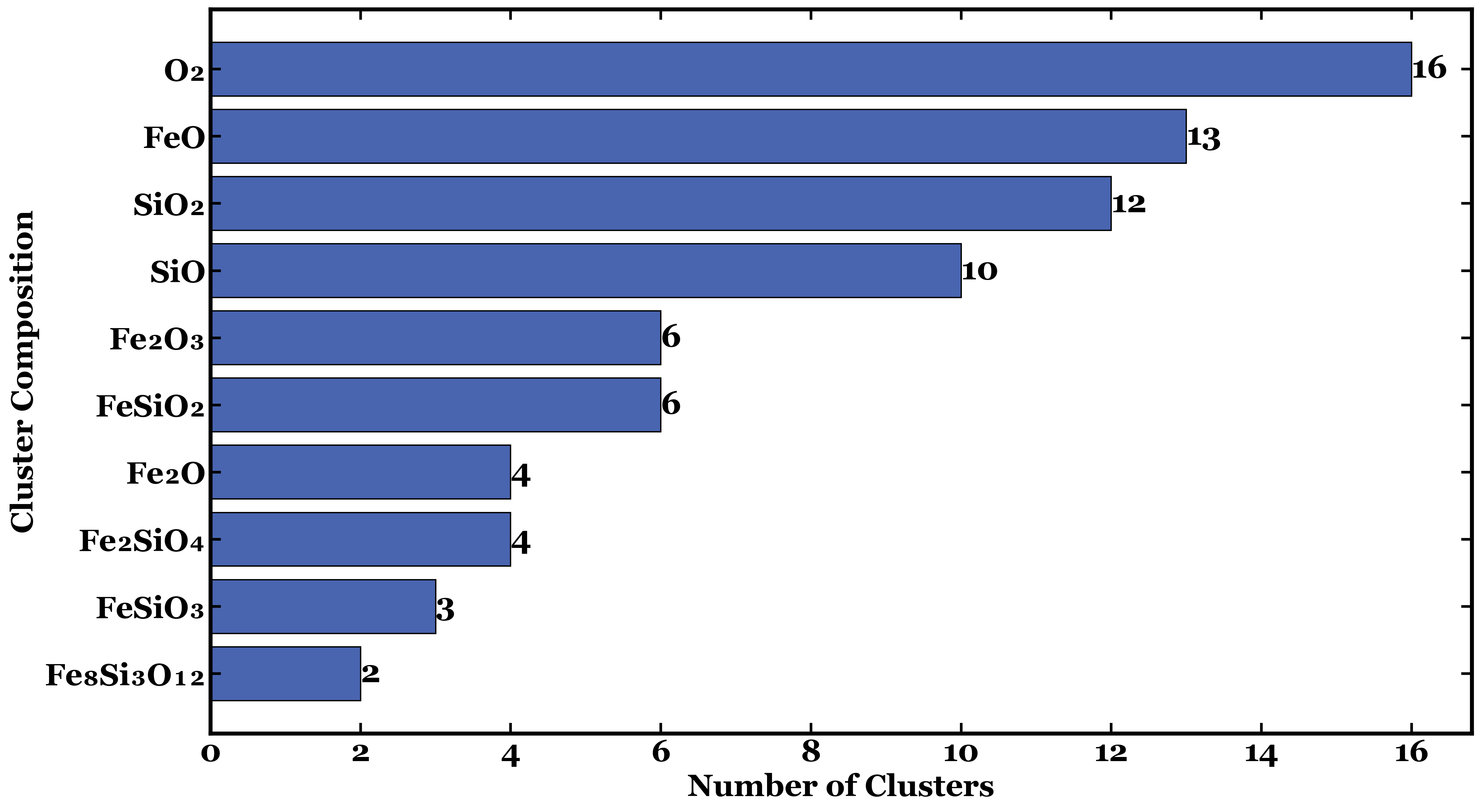}%
    \caption{Ten most frequently observed elemental compositions of clusters ejected during the impact event. Each horizontal bar represents a unique cluster composition identified through spatial clustering and subsequent elemental analysis, with the length of the bar directly correlating to the number of clusters found for that specific composition.
    The compositions are ordered from most to least frequent, allowing for immediate identification of the dominant ejected species.}
    \label{fig:VDF_ejecta}
\end{figure}

We then focused on the ejecta and analyzed the chemical dynamics occurring during the impact event by identifying and quantifying the elemental composition of these clusters. Figure~\ref{fig:Ejecta_Comp} (a) to Figure~\ref{fig:Ejecta_Comp} (f) shows the chemicals found in the fully expanded plume by cluster analysis and the statistics of the clusters is plotted in Figure~\ref{fig:VDF_ejecta}. Exploring the resulting ejecta gives insight into the primary mechanisms governing material ejection and potential chemical transformations.  
 The prevalence of specific cluster types provides a fingerprint of the altered environment and the stability of various chemical species under extreme conditions. The most interesting finding from the cluster composition analysis is the dominance of O$_2$, which stands out as the single most frequent ejected species. This high abundance of diatomic oxygen in the plume strongly indicates the formation of an oxygen-rich environment immediately following the impact. This suggests significant bond breaking within the original material and the subsequent recombination of highly mobile oxygen atoms into stable molecular units, which are then ejected.

Beyond diatomic oxygen, the presence of FeO and SiO$_2$ as common ejecta fragments points to the direct decomposition of the parent material, Fe$_2$SiO$_4$ (fayalite). These compositions suggest that during fragmentation, iron and silicon largely maintain their typical oxidation states, often accompanied by oxygen in stoichiometric ratios indicative of their oxide forms. 
 The emergence of SiO implies a lower oxidation state for silicon than in SiO$_2$, {\color{black}
which indicates that our calculated Fe oxidation states likely represent lower-limit estimates, and the actual extent of Fe$^{3+}$ formation may be slightly underestimated.} 
while Fe$_2$O indicates a reduction of iron, possibly from its typical +2 state in olivine to a +1 oxidation state in these ejected fragments.
{\color{black}
The FeSiO$_2$ clusters observed in the fully expanded plume do not correspond to known stable phases but likely represent transient coordination molecules formed under nonequilibrium conditions.
}
The continued presence of Fe$_2$SiO$_4$ itself also demonstrates that some portions of the original material either remain intact or re-form following the high-energy impact.

\subsection{Oxidation State}

\begin{figure}[!h]
    \centering
    \includegraphics[width=1.0\textwidth]{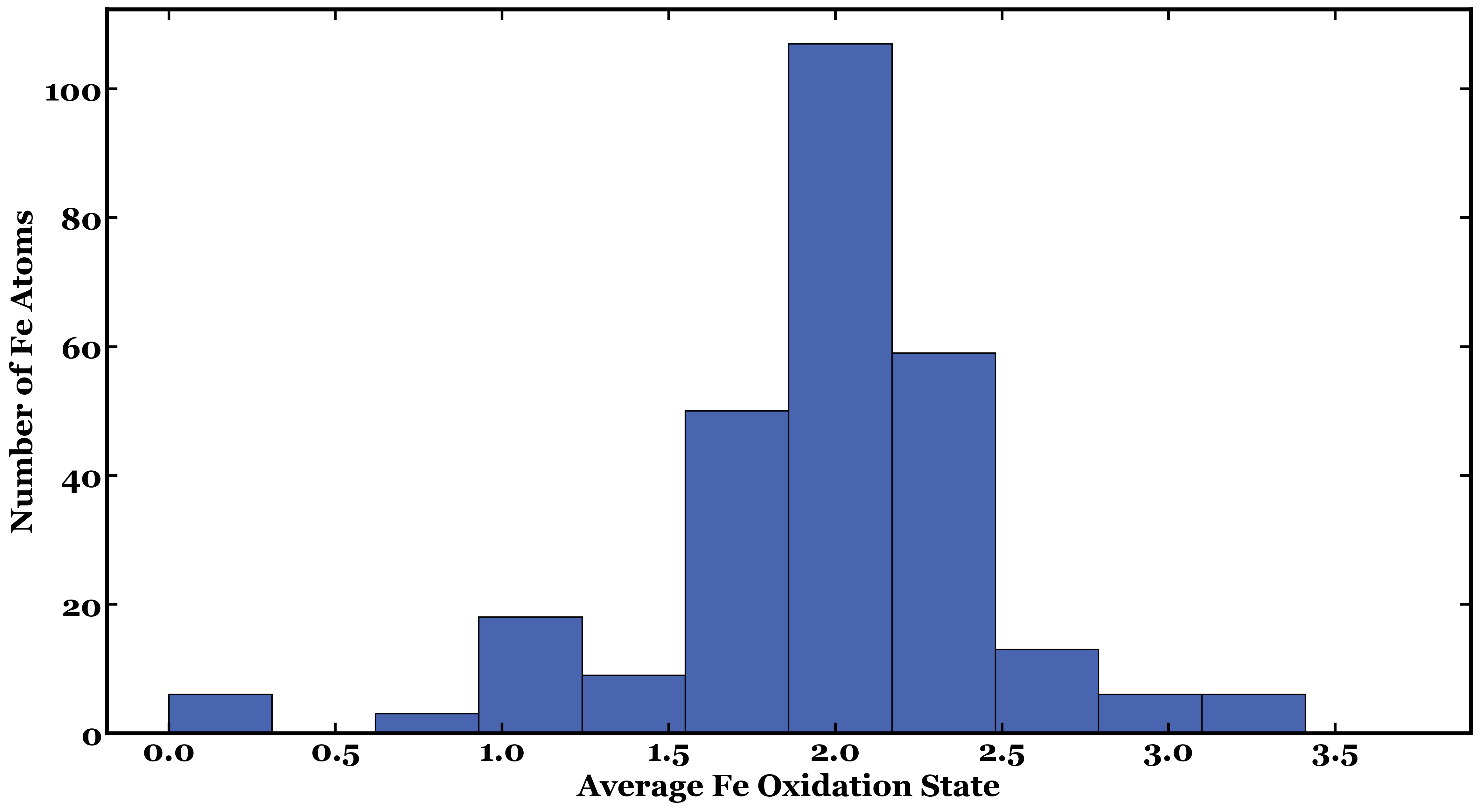}%
    \caption{Oxidation state distribution of iron in impact ejecta. By analyzing the composition of ejected species such as Fe$_2$SiO$_4$ and assuming fixed oxidation states for O (-2) and Si (+4), the oxidation state of Fe was calculated. }
    \label{fig:hist_oxidation}
\end{figure}

The distribution of iron's oxidation states within the impact ejecta, as depicted in Figure \ref{fig:hist_oxidation}, provides critical insights into the redox conditions prevailing during and immediately after the high-energy event. By analyzing the elemental composition of ejected species such as Fe$_2$SiO$_4$ (olivine) and assuming fixed oxidation states for oxygen (O$^{-2}$) and silicon (Si$^{+4}$), we calculated the average oxidation state for iron within each identified cluster. The results indicate that Fe(II) (Fe$^{2+}$) is the predominant oxidation state, reflecting its abundance in the parent olivine mineral. The histogram shows a dominant peak centered at an oxidation state of +2.0, with significant, roughly symmetrical distributions extending towards both lower values (down to approximately +1.5) and higher values (up to approximately +2.5). This corresponds to the expected prevalence of ferrous iron (Fe$^{2+}$) from the parent olivine, but critically, also indicates the occurrence of both reduction (leading to species with average Fe charge between +1.5 and +2.0) and oxidation (leading to species with average Fe charge between +2.0 and +2.5) processes within the ejecta.

Further analysis of Figure~\ref{fig:hist_oxidation} reveals the presence of clusters containing Fe--Fe bonds, with the largest continuous Fe--Fe network comprising up to three iron atoms (Fe$_3$) within the largest cluster (Fe$_{57}$O$_{102}$Si$_{24}$). The detection of these reduced iron species---manifested as direct Fe--Fe connectivity---suggests that localized chemical environments within the system can facilitate partial reduction of iron, enabling the formation of metallic iron clusters at the nanoscale (npFe$^0$) even in the absence of bulk metallic phases. In contrast, the observation of clusters with average iron oxidation states extending toward $+2.5$, and the broader distribution around the Fe$^{2+}$ peak, indicates that other regions favor the stabilization of more oxidized iron species, likely influenced by the local availability of oxygen and distinct reaction pathways during rapid cooling.  It should be noted that reduction of Si is also observed in the ejecting plume, and we assume Si as a more stable element remain its oxidation state (+4), but it's also possible those Si within iron-silicon-bearing clusters are also reduced leading to even higher oxidation state of Fe.  Overall, the distribution of Fe oxidation states serves as a sensitive indicator of the complex chemical processes and kinetic effects that occur during the evolution of the ejecta plume.

\subsection{Disproportionation Reactions}

Our simulations also offer a dynamic perspective distinct from purely stoichiometric reactions like disproportionation ($3\text{Fe}^{2+} \rightarrow \text{Fe}^0 + 2\text{Fe}^{3+}$). While disproportionation undoubtedly contributes to lunar iron chemistry, the hypervelocity impact environment involves a cascade of bond-breaking, atomic mixing, and rapid quenching. The observed Fe$^{3+}$ in ejecta arises from these complex, non-equilibrium dynamics, which define immediate reaction pathways and species formation that are difficult to predict without explicitly modeling the atomic-scale interactions and energy transfers inherent to such energetic events.

Furthermore, while our simulations robustly show the formation of Fe$^{3+}$ species, particularly within the ejecta plume, at the impact site, the direct formation of substantial crystalline Fe$_2$O$_3$ (hematite) is less prevalent within the short timescales simulated. Instead, we observe a significant abundance of Fe-Fe bonds, indicative of nanophase metallic iron (npFe$^0$) formation, especially in regions highly affected by impact cratering. This suggests that while impacts readily generate Fe$^{3+}$ precursors, their subsequent conversion to hematite likely requires prolonged interaction with oxidants and favorable conditions post-impact, rather than being an immediate, widespread product of the impact itself. This offers an interesting perspective on the effective disproportionation induced by micrometeoroid impacts, where the immediate impact-processed surface is left in a more reduced state, consistent with observations of npFe$^0$ in Apollo and Chang'E mission samples, while the ejected plume carries more oxidized species.

\subsection{Evolution of Nanophase Iron and Implications of Mixed Oxidation States}

Our findings of impact-induced iron disproportionation align with and provide a potential initial mechanism for observations made by \cite{thompson2016oxidation}. Their electron energy loss spectroscopy (EELS) studies on lunar soil nanoparticles revealed mixtures of \textit{$Fe^0$, $Fe^{2+}$, and $Fe^{3+}$} states, with an overall oxidation increase in more mature soils, potentially due to oxygen diffusion. The impact-generated $Fe^0$ at the impact site and $Fe^{3+}$ in the ejecta would then undergo these longer-term maturation processes. For instance, the metallic iron could gradually oxidize over time, while the ejected $Fe^{3+}$, already oxidized, might transform mineralogically, perhaps crystallizing into ferric oxides depending on post-depositional conditions. This offers a critical link between immediate impact chemistry and regolith evolution.

Standard optical maturity indices for lunar soils are often linked to the abundance and characteristics of np$Fe^0$. If impact-induced disproportionation is a significant contributor to the np$Fe^0$ population, and if its efficiency varies with impactor energy, velocity, or target mineralogy (e.g., olivine vs. pyroxene vs. ilmenite content), this could introduce additional complexities in the interpretation of maturity based solely on these indices. The presence of different generations or types of np$Fe^0$ (e.g., from solar wind reduction versus impact disproportionation) might need to be considered for a more nuanced understanding of regolith evolution. Understanding this impact-driven redox differentiation is therefore crucial for accurately modeling the optical properties of the lunar surface. It also bears significantly on the interpretation of compositional data from past, present, and future lunar missions, including orbital remote sensing and analyses of returned samples, as the redox state of iron profoundly influences its spectral behavior, magnetic properties, and chemical reactivity.
  
\subsection{Implications for Hemitate Observed by M$^3$}

\cite{Li2020} observed hematite predominantly at high latitudes, often associated with east- and equator-facing sides of topographic highs and more prevalent on the nearside. Their leading hypothesis for its formation is the oxidation of lunar surface iron (Fe(II) or Fe$^{0}$) by oxygen delivered from Earth's upper atmosphere, particularly when the Moon is within Earth's magnetotail, a period offering shielding from the reducing solar wind. They further suggest that lunar water (H$_2$O/OH), more abundant at these high latitudes, plays a crucial role in facilitating these oxidation reactions. Micrometeoroid impacts are considered by \cite{Li2020} as agents that can enhance hematite formation by increasing reaction kinetics, gardening the surface, and liberating water, but not as the primary source of the oxidized iron via disproportionation. Our MD simulation results propose an alternative, or perhaps complementary, pathway for the initial generation of Fe(III) species directly from lunar materials during impact events. 
Our model proposes an intrinsic process where impact energy drives the disproportionation of lunar Fe(II), creating Fe(III) without requiring an external oxygen source for this initial oxidation step. 

In addition, \cite{Li2020} reported that hematite occurrences, while much less frequent, are not entirely absent on the farside of the Moon, where the delivery of oxygen from Earth is unlikely. This observation highlights a key challenge in fully explaining the strong dichotomous distribution of hematite between the near and far sides of the Moon, where hematite is predominantly found on the nearside. The impact-induced disproportionation mechanism proposed here, being intrinsic to the physical and chemical conditions of the impact process, offers a plausible explanation for the presence of Fe(III) minerals in these farside regions or in areas shielded from Earth’s influence, where models invoking Earth-derived oxygen are less applicable.

Considering the effect of Earth-derived O$^{+}$   in nearside ejecta plumes could indeed show enhanced oxidation processes, thereby providing a more comprehensive explanation for the observed near-far side dichotomy in hematite distribution. This suggests that while Earth-derived oxygen may play a significant role on the nearside, intrinsic lunar processes like impact disproportionation are crucial for understanding farside hematite occurrences. Furthermore, if impact-driven disproportionation was an active process throughout lunar history, it might have contributed to the localized presence of Fe(III)-bearing phases even before Earth's atmosphere became significantly oxygenated or before Earth-Moon atmospheric transfer dynamics were established as they are today, allowing for global oxidation even in the absence of significant Earth-Moon atmospheric transfer.

\section{Conclusion}

Our molecular dynamics simulations using the ReaxFF reactive force field provide valuable insights into the complex chemical dynamics triggered by micrometeoroid impacts on lunar surfaces. Echoing our introduction, this study underscores that impacts can simultaneously generate spatially distinct reducing and oxidizing environments, directly influencing lunar regolith chemistry and space weathering processes. Within the high-pressure, high-temperature crater environment, the formation of nanophase metallic iron (npFe$^0$) predominates, confirming impact-induced reduction pathways crucial to understanding lunar surface maturation.

Conversely, our analyses of the ejecta plume revealed significant oxygen mobilization and the formation of oxidized iron species, notably Fe$^{3+}$-bearing clusters. These results highlight the potential for localized oxidation under specific impact-driven conditions, aligning with observations of hematite in lunar polar regions reported by M$^3$. Although direct crystalline hematite formation was not prevalent within our simulation's limited timescales, the presence of oxidized iron precursors suggests a feasible pathway for hematite formation upon subsequent exposure to environmental oxidants.

{\color{black}Previous studies, such as \citep{li2024vacuum}, investigated thermal alteration of troilite (FeS) under thermal conditions using ab initio deep neural network models and thermodynamic calculations. In contrast, our simulations employ a reactive molecular dynamics framework (ReaxFF) to directly model hypervelocity impacts on Fe-rich silicate minerals. This allows us to distinguish the different redox conditions on the surface and in the ejecta plume within a single simulation, offering a mechanistic explanation for the co-occurrence of npFe$^0$ and hematite on the Moon.}

Critically, this work illustrates the dynamic and transient nature of impact-induced redox processes, highlighting the inadequacy of equilibrium-based or static models alone to describe the full scope of lunar space weathering. By elucidating these atomistic pathways, our findings bridge previously disparate observations—linking impact-driven npFe$^0$ formation to the puzzling presence of oxidized iron phases on the Moon. 
{\color{black}
In this study, we adopted a fixed impact velocity of 12 km/s and a 45° incidence angle, which together represent typical micrometeoroid conditions on the Moon (Grün et al., 2011; Melosh, 1989). While variations in velocity and angle may influence the details of plume morphology and chemical yields—especially for volatile species—we expect the key redox mechanism identified here, involving spatial separation between reduced and oxidized iron, to remain qualitatively valid. This mechanism is primarily driven by energy density and atomic-scale dynamics rather than geometric symmetry.} {\color{black}While our stoichiometry-based oxidation state estimates provide useful insight into redox trends, determining the exact oxidation states of iron—especially in complex plume species—will require future work using ab initio molecular dynamics or improved reactive potentials that more accurately capture charge transfer in mineral systems.}
Future research incorporating longer simulation times and 
{\color{black}changes in impact parameters}
will further refine our understanding of these intriguing lunar redox processes.

\begin{acknowledgments}

We acknowledge useful feedback and discussions from Brant M. Jones and Shuai Li. This research was supported in part through research cyber infrastructure resources and services provided by the Partnership for an Advanced Computing Environment (PACE) at the Georgia Institute of Technology, Atlanta, Georgia, USA. Z.H., M.H. and T.M.O are supported by SSERVI/CLEVER (NNH22ZDA020C/80NSSC23M022). M.H. acknowledges support by VIPER (80NSSC24K0682) and SSERVI/RASSLE (80NSSC24M0016).

\end{acknowledgments}

\begin{contribution}
 
Z.H came up with the initial research concept and was responsible for writing and submitting the manuscript. Z.H provided the formal analysis and validation. M.H and T.M.O obtained the funding and edited the manuscript. All authors help with writing and editing the manuscript.

\end{contribution}
 
\software{scipy \cite{virtanen2020scipy}, LAMMPS \cite{thompson2022lammps}, OVITO \cite{stukowski2009visualization}
          }

\bibliography{sample7}{}
\bibliographystyle{aasjournalv7}

\end{document}